\documentclass[aps,prl,twocolumn,amsmath,amssymb, superscriptaddress]{revtex4-2}
\usepackage{graphicx}% Include figure files
\graphicspath{{Figures/}}
\usepackage{subfigure}
\usepackage{epsfig}
\usepackage{ulem}
\usepackage{dcolumn}% Align table columns on decimal point
\usepackage{bm}% bold math
\usepackage{hyperref}% add hypertext capabilities
\hypersetup{colorlinks=true, citecolor=blue, urlcolor=blue, linkcolor=blue}
\newcommand{\Tr}{{\rm Tr}}

\newcommand{\ra}{\rightarrow}

\def\beq{\begin{equation}}
\def\eeq{\end{equation}}
\def\bald{\begin{aligned}}
\def\eald{\end{aligned}}
\def\bea{\begin{eqnarray}}
\def\eea{\end{eqnarray}}

\def\ket#1{\left|#1\right\rangle}
\def\avg#1{\left\langle#1\right\rangle}

\def\expectation#1#2#3{\left\langle #1\left| #2 \right| #3\right\rangle}
\def\Tr{\mathrm{Tr}}

\def\Eq#1{Eq.~(\ref{#1})}
\def\Fig#1{Fig.~\ref{#1}}

\begin{document}
\title{Asymptotic sign free in interacting fermion models}
\author{Zi-Xiang Li}
\thanks{These authors contributed equally to the work.}
\affiliation{Beijing National Laboratory for Condensed Matter Physics and Institute of Physics,
Chinese Academy of Sciences, Beijing 100190, China}
\affiliation{University of Chinese Academy of Sciences, Beijing 100049, China}
\author{Zhou-Quan Wan}
\thanks{These authors contributed equally to the work.}
\affiliation{Institute for Advanced Study, Tsinghua University, Beijing 100084, China.}
\author{Hong Yao}
\email{yaohong@tsinghua.edu.cn}
\affiliation{Institute for Advanced Study, Tsinghua University, Beijing 100084, China.}

\begin{abstract}
As an intrinsically-unbiased approach, quantum Monte Carlo (QMC) is of vital importance in understanding correlated phases of matter.
Unfortunately, it often suffers notorious sign problem when simulating interacting fermion models. 
Here, we show for the first time that there exist interacting fermion models whose sign problem becomes less severe for larger system sizes and eventually disappears in the thermodynamic limit, which we dub as ``asymptotic sign free''. 
We demonstrate asymptotically-free sign in determinant QMC for various interacting models.
Moreover, based on renormalization-group-like ideas we propose a heuristic understanding of the feature of asymptotic sign free.
We believe that asymptotic sign free behavior could shed new lights to deepening our understanding of sign problem. More importantly, it can provide a promising way to decipher intriguing physics in correlated models which were conventionally thought not accessible by QMC.
\end{abstract}
\date{\today}

\maketitle
{\bf Introduction:} Understanding novel correlation physics such as high-temperature superconductivity \cite{Keimer2015Nature, Davis2013, Lee2006RMP, Scalapino-RMP2012, Fradkin2015RMP} in interacting systems has been one of central issues in modern condensed matter physics and other related fields. 
However, interacting quantum many-body systems in more than one dimension, especially those with strong correlations, are generally beyond the solvability of analytical approaches with well theoretical control.
Therefore, developing efficient numerical approaches to study quantum correlated systems is thus of vital importance.
Quantum Monte Carlo (QMC) \cite{ZXLiQMCreview,Assaad2008world, Suzuki1993Book, Scalapino1981PRL, Blankenbecler1981PRD, Hirsch1981PRL,Hirsch1983PRB, White1989PRL,Zhang1995PRL, Gull2011RMP,Sandvik1991PRB,Sandvik2002PRE,Kaul2013ARCMP}, which simulates quantum many-body systems by stochastic sampling in an intrinsically-unbiased way, is among the most important approaches. 
Unfortunately, QMC simulations of quantum models often suffer from notorious sign problem, namely the weights of sampling configurations may not be positive definite for fermionic models \cite{Loh1990PRB,Scalettar1990PRB} and frustrated spin models \cite{Hatano1992PLA, Sandvik2000PRB, Nakamura1998PRB}. Sign problem is currently the main obstacle in applying QMC to study quantum many-body physics efficiently. It is then highly desired to investigate general features of sign problem and to solve sign problem of various interacting models potentially hosting intriguing physics. %such as high-temperature superconductivity.

\begin{figure}[t]
\includegraphics[scale=0.38]{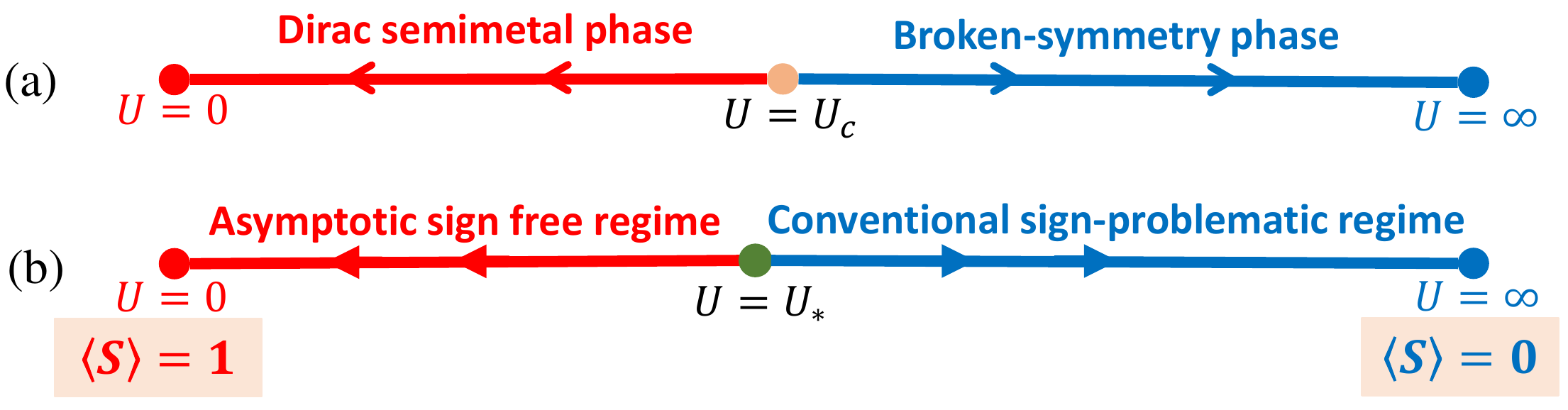}
\caption{The schematic representation of the RG flow of interaction $U$ in (a) and the RG-like flow of average sign $\avg{S}$ in (b). (a) There is a physical quantum phase transition at critical interaction $U=U_c$ between two distinct phases (one at weak coupling and the other at strong coupling). (b) There is a sign transition point at $U=U_*$ between two distinct regimes of sign scaling behaviors. When the interaction is relatively weak ($U<U_*$), the average sign $\left<S\right>$ increases asymptotically to one as the system size increases, which is qualitatively different from the conventional exponential decaying behaviour of the average sign for strong interactions ($U>U_*$). Note that the physical phase transition point $U_c$ is close to the sign transition point $U_*$ although they are in general different.}
\label{Fig1}
\end{figure}

Even though it was shown that a generic solution to the sign problem is nondeterministic polynomial (NP) hard \cite{Troyer2005PRL}, sign problem can be cured for many specific quantum models. In the past two decades, tremendous efforts have been made in solving sign problems by various approaches, including designing new algorithms \cite{Chandrasekharan2016PRB, Chandrasekharan1999PRL},  utilizing complex-fermion symmetries   \cite{Wu2005PRB,Berg2012Science}, employing Majorana representations \cite{Li2015PRB,Li2016PRL}, and  mathematical structures \cite{Zhang2003PRL, Xiang2016PRL, Wang2015PRL} (for a recent review, see, e.g., Ref. \cite{ZXLiQMCreview} by two of us).
Indeed, various intriguing physics has been revealed in studying strongly correlated models by sign-free QMC simulation \cite{Li2015NJP, Wang2014NJP, Li2017NC, Xu2017PRX, Kivelson2016PRX, Li2018ScienceAdvances, Li2017PRL, Berg2019Review, Assaad2013PRX, Kivelson2018PRB, Scalettar2021PRL, Sachdev2014NP, Sachdev2018PNAS,
Vishwanath2016PRB, Wu2011PRB, Assaad2011PRL, Assaad2005PRB, Xu2019PRX, Grover2018PRL, Lang2019PRL, Berg2016PRL, Trebst2007PRL, Sandvik2007, Shao2016Science, Adam2018Science, Sandvik2010Review, Xu2021PRL, Vishvanath2017NP, Berg2021arXiv}. Additionally, new strategies such as machine learning are developed to mitigate sign problem \cite{Broecker2017,wan2020mitigating, Hangleiter2020ScienceAdvances, Sorella2007PRL, Mila2020PRB, Abolhassan2021PRL} in QMC simulations.
When sign problem does appear in QMC simulation of a model, %namely weight of configuration is not positive definite,
the average sign usually decays exponentially with system size and inverse temperature \cite{Troyer2005PRL} such that the needed computational time increases exponentially with system size and inverse temperature, hampering reliable QMC simulations for large system size and low temperature. 

Here, we demonstrate a novel behavior of sign problem, which is qualitatively distinct from conventional sign problem mentioned above.
For the first time, we numerically observed that the sign problem of various interacting models are increasingly mitigated when their system size is increased, although the sign-problem of those models cannot be fully solved by any known methods so far.
Specifically, the average value of sign can approach asymptotically to one, namely sign problem in the simulation asymptotically vanishes, in the limit of infinite system size.
We dub this intriguing phenomenon as ``asymptotic sign-free'' (ASF). Our study of ASF mainly focuses on Dirac fermion systems, namely models whose non-interacting limit features Dirac cones, although we will discuss other types of models as well.

To understand this novel observation, we propose a renormalization group (RG) like explanation of ASF, which might provide a possible general theoretical framework to fathom the asymptotic behavior of sign problem in interacting systems.
The spirit of RG-like scenario to understand the scaling behaviour of sign problem in interacting models is illustrated in \Fig{Fig1}.
We believe that the asymptotic sign free behavior unveiled in our study will shed new insight to understanding the nature of sign problem, and potentially paves a promising avenue to studying intriguing physics in correlated fermion models with large size which were previously thought not accessible by QMC.

{\bf Sign problem in QMC:} In this paper we employ determinant QMC (DQMC), which is an intrinsically-unbiased QMC algorithm, to simulate interacting fermion models.
Both finite-temperature and ground-state properties of a model can be investigated by DQMC.
For finite-temperature simulations, the expectation value of a physical observable can be computed as:  $\langle\hat{O}\rangle=\frac{\Tr{e^{-\beta \hat{H}}\hat{O}}}{\Tr{e^{-\beta \hat{H}}}}$, where $\beta$ is the inverse temperature and $\hat H$ is the Hamiltonian.
Employing Hubbard-Stratonovich (HS) transformation of the interacting terms in the Hamiltonian which introduces auxiliary fields, the expectation value can be expressed: $\langle \hat{O} \rangle= \frac{\sum_{z} O(z) w(z)}{\sum_{z} w(z)}$, where $O(z)$ is the expectation value of observable $\hat{O}$ for auxiliary field configuration $z$ and $w(z)$ %
%= \Tr \prod_{l=1}^{N_\tau} e^{-\hat{H}_0 \Delta_\tau} e^{-\hat{H}_s \Delta\tau }$
is the Boltzmann weight determining the sampling probability in simulations.

The sign problem appears once the weight $w(z)$ is not positive definite. When the sign $S(z)\!=\!\frac{w(z)}{|w(z)|}$ can be positive or negative, its average value $\langle S \rangle =\frac{\sum_z w(z)}{\sum_z |w(z)|}$ for a sign-problematic model usually decays exponentally with system size $N$ and inversion temperature $\beta$, $\langle S\rangle \sim e^{-\kappa \beta N}$ ($\kappa>0$) \cite{Troyer2005PRL}, which renders QMC simulations not feasible for large system size and low temperature.
In the following, we perform large-scale QMC simulations
on several quantum many-body models featuring Dirac fermions to demonstrate that asymptotic sign free behavior can emerge, namely the average sign $\langle S\rangle$ increases with system size, rendering reliable QMC studies of those models with large system size feasible although those models are sign-problematic.

\begin{figure}[t]
\includegraphics[scale=0.4]{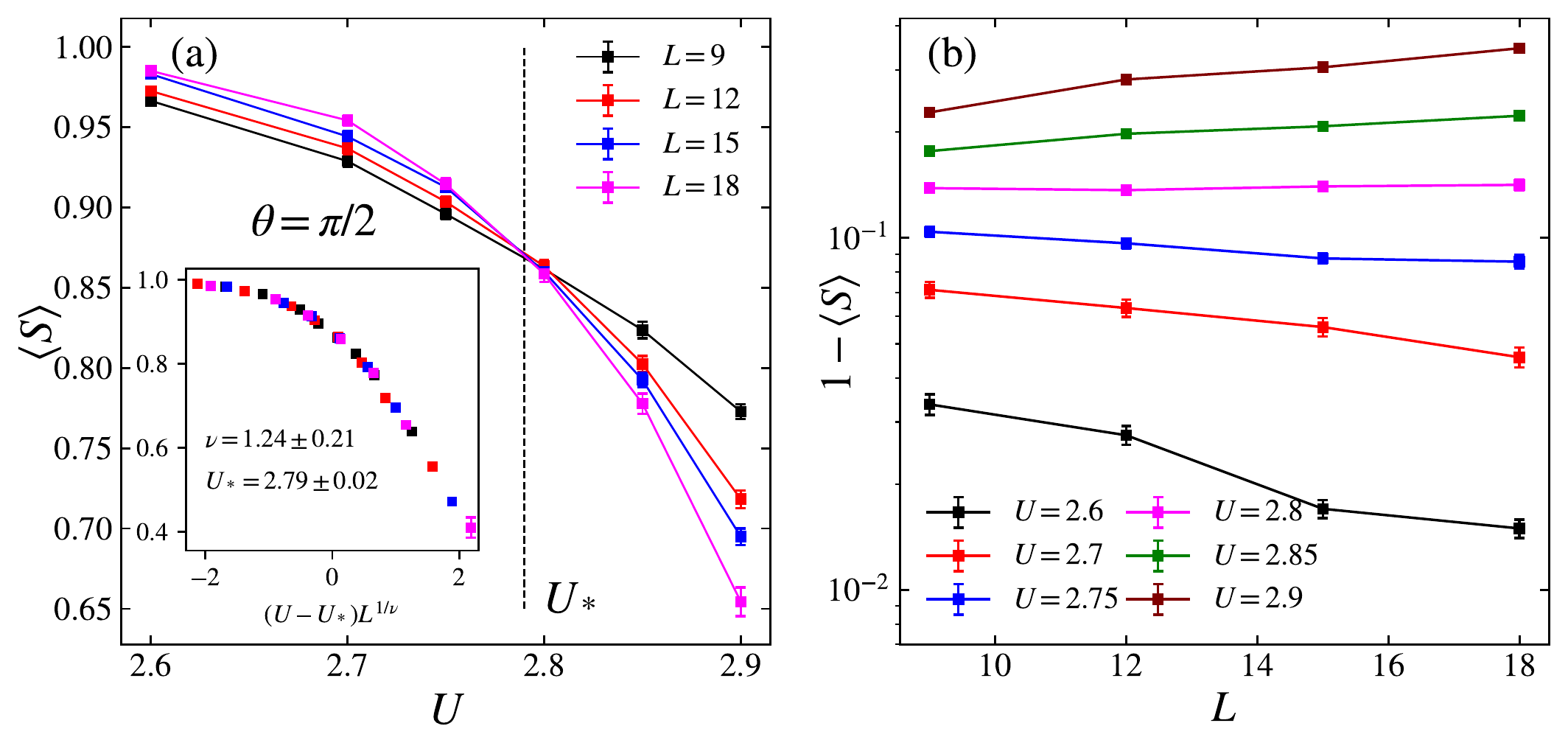}
\caption{The average sign $\left<S\right>$ in QMC simulations of the honeycomb Hubbard model at half-filling using a special type of Hubbard-Stratonovich transformation. (a) $\left<S\right>$ as a function of interaction strength $U$. Here $U=U_*$ represents a transition between sign scaling behaviour. For $U<U_*$, the sign problem becomes less severe for larger system while for $U>U_*$ opposite behaviours occurs. Inset is the result of data collapse using $\left<S\right>$ and $(U- U_*)L^{1/\nu}$. (b) The scaling of $1-\left<S\right>$ as a function of linear system size $L$. The calculation is performed for linear system size $L=9,12,15,18$ with inverse temperature $\beta = 2 L$. For $U<U_*$, the average sign $\left<S\right>\approx 1-e^{-\alpha L}$ ($\alpha>0$) and it approaches to 1 in the thermodynamic limit, namely asymptotic sign free.}
\label{Fig4}
\end{figure}

{\bf The honeycomb Hubbard model:} The interacting effect on the Dirac fermions has stimulated enormous interests in the past many years, including interaction-driven quantum phase transitions \cite{Li2015NJP, Wang2014NJP, Li2017NC, Jian2017PRB, Herbut2006PRL, CJWu-PRB2018, Tang2018Science, Assaad2017PRL, Herbut2016PRB, Assaad2015PRB, Sarma2014PRB, Herbut2013PRX, Lu2012PRB} and topological phases \cite{Qi2011RMP, Kane2010RMP, Wu2011PRB, Assaad2011PRL, Raghu2008PRL, Honerkamp2015PRB, Vishwanath2012PRL, Herbut2013PRB, Franz2010PRB, Li2017PRB}. One simple but important model featuring interacting Dirac fermions is the spin-1/2 Hubbard model on the honeycomb lattice. We now consider this model:
\bea\label{Model2}
H=-t\sum_{\avg{ij},\sigma}(c^\dagger_{i\sigma}c_{j\sigma}+h.c.) + U\sum_{i} n_{i\uparrow}n_{i\downarrow},
\eea
$c^\dag_{i\sigma}$ creates an electron on site $i$ with spin polarization $\sigma=\uparrow$/$\downarrow$, $t$ is the NN hopping, and $U$ is the onsite repulsion. Hereafter we set $t=1$ as unit of energy.
The honeycomb Hubbard model at half filling is sign free when choosing appropriate HS transformations and has been shown to exhibit a quantum phase transition at $U\!=\!U_c\!\approx\! 3.85$ between Dirac semimetal and antiferromagnetic Mott insulator \cite{sorella2016PRX}.
Here we employ a special HS transformation to purposely make sign problem occur in this model by decoupling the Hubbard interaction differently between terms on A sublattices and B sublattices. 
Specifically, we decouple the Hubbard interaction in the spin channel as follows: $
e^{-\Delta \tau U(n_{i\uparrow}-\frac 1 2)(n_{i\downarrow}-\frac 1 2)}=\frac{1}{2} e^{-U\Delta \tau/4}\sum_{s_i =\pm 1}e^{\lambda s_i c^\dagger_i \vec{\sigma} \cdot \vec{n}_i c_i}$,
where $c^\dag_i=(c^\dag_{i\uparrow},c^\dag_{i\downarrow})$, $\Delta\tau$ is the time-slice in Trotter decomposition, $\lambda$ is a constant defined as $\cosh \lambda=\exp(U\Delta\tau/2)$, $s_i$ is axillary field, $\vec{\sigma}=(\sigma^x,\sigma^y,\sigma^z)$ are Pauli matrices, and $\vec n_i$ is a unit vector.
Specifically, we set $\vec n_i=\hat z$ for $i\in$ A sublattice and $\vec n_i=\vec n =(\sin \theta, 0, \cos\theta)$ for $i\in$ B sublattice. %the Hubbard interaction is decoupled in $s_z$ channel while for the B sublattice it is decoupled in a rotated spin channel as follows:
%\bea\label{AB_HS}
%e^{-\Delta \tau U(n_\uparrow-\frac 1 2)(n_\downarrow-\frac 1 2)}=\frac 1 2e^{-U\Delta \tau/4}\sum_{s =\pm 1}e^{\lambda s c^\dagger \mathbf{\sigma} \cdot \mathbf{n}c},
%\eea
%where $\cosh \lambda=\exp(U\Delta\tau/2) $ and $\mathbf{n}=(\sin \theta,0,\cos \theta)$.
When $\theta =0$, this HS transformation becomes the conventional one. 
When $\theta \neq 0$, QMC simulations will suffer a sign problem and the severity of sign problem can be continuously tuned by varying $\theta$.
Using this special HS transformation with fixed $\theta$, we investigate how the sign problem of this model depend on the system size under study.

For $\theta=\pi/2$, curves of the average sign $\langle S \rangle$ for QMC simulations of the model on the lattice with $L\!\times\! L$ unit cells show crossing behavior around $U=U_*\approx 2.79$, as shown in \Fig{Fig4}(a).
The crossing suggests that $U=U_*$ represents a transition between different sign scaling behaviours.
For the sign transition at $U=U_*$, we obtain the nominal critical exponent $\nu_*\approx 1.2$, which appears to be close to the physical correlation-length exponent $\nu \approx 1.02$ of the physical quantum phase transition at $U=U_c$ \cite{sorella2016PRX}.
Remarkably, the average sign $\langle S \rangle$ in the relatively weak interaction regime $U<U_*$ increases with the system size $L$ with an approximate scaling $\avg{S}\approx 1-e^{-\alpha L}$ ($\alpha>0$) as shown in \Fig{Fig4}(b). 
Consequently, the average sign flows to one in the thermodynamic limit, namely asymptotic sign free. Conversely, when Hubbard interaction is relatively strong, namely $U>U_*$, the average sign decays with system size $L$ and the model exhibits conventional sign problematic behaviour.
We further calculate the average sign in QMC simulations of this model using different values of $\theta$ in the HS transformation, as shown in \Fig{FigS1}; similar crossing occurs at a slightly different value of $U$ but the asymptotic sign free behavior is robust in the weak interaction regime.

\begin{figure}[t]
\includegraphics[scale=0.4]{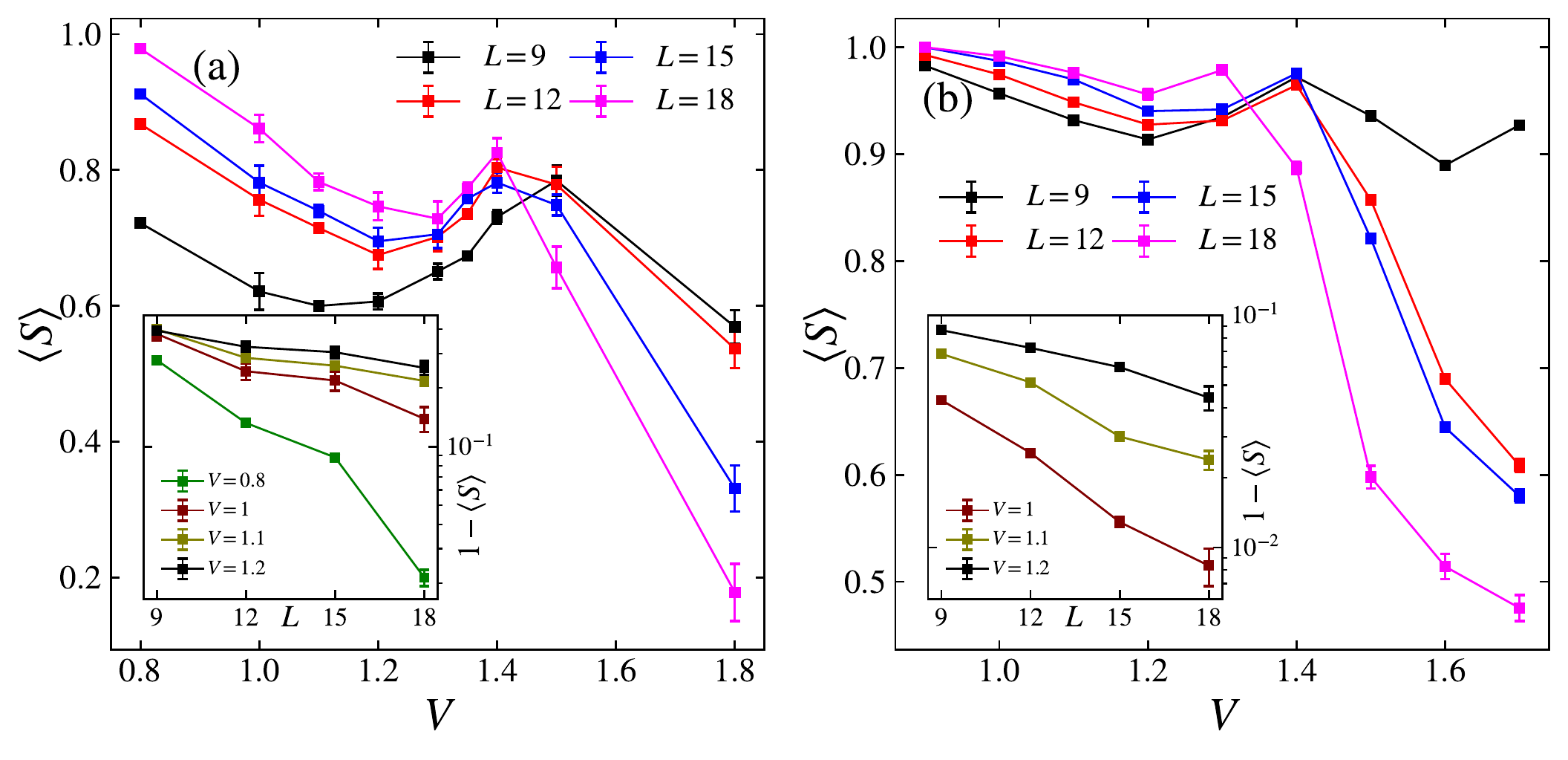}
\caption{(a) The results of the average sign $\avg{S}$ in zero-temperature QMC for the spinless fermionic model with repulsive interaction $V>0$. The calculation is performed for linear system size $L=9,12,15,18$. The projector parameter is fixed $\Theta = 60$.  (b) The results of the average sign $\avg{S}$ in finite-temperature QMC for the same model as in (a). The calculation is performed for $L=9,12,15,18$. The inverse temperature $\beta$ scales with $L$ as $\beta = 3 L$. The insets explicitly show the results of average sign $\avg{S}$ versus $L$ for several values of $V$ in the weak interaction regime.  }
\label{Fig2}
\end{figure}

{\bf Repulsive spinless honeycomb model:} In addition to the spinful Hubbard model, we investigated the spinless fermion model on honeycomb lattice with the following Hamiltonian:
\bea\label{Model1}
H=-t\sum_{\avg{ij}}(c^\dagger_{i}c_{j}+h.c.) + V \sum_{\avg{ij}} (n_i-\frac{1}{2})(n_j-\frac{1}{2}),
\eea
where $c_i$ annihilates a spinless fermion on site $i$, $n_i = c^\dagger_ic_i$ is the fermion number operator, $t$ is the hopping amplitude of fermions on the NN bond, and $V>0$ represents density repulsion. 
When $V=0$, the model in \Eq{Model1} exhibits massless Dirac fermions. We focus on half filling, namely Fermi level lies at the Dirac point.

To perform QMC study of the model above, one needs to do HS transformation to decouple the density-density interaction and there are various ways to do it (details of the HS transformation in different channels are shown in the Supplementary Materials).
If the interaction term is decoupled in the density channel, the QMC simulations displays sign problem; namely the weight $w(z)$ is not positive definite. 
It was known that there is a quantum phase transition at $V=V_c$ between the Dirac semimetal at weak interactions and CDW ordering at strong repulsions \cite{Wang2014NJP, Li2015NJP, Wessel2016PRB, Chandrasekharan2017PRD, Wang2016PRB}.
To investigate the nature of sign problem of this model, we calculate the average sign in the sign-problematic decoupling channel, and explore how the behavior of sign problem depends on interaction strength $V$ and on system size $L$.

\begin{figure}[t]
\includegraphics[scale=0.4]{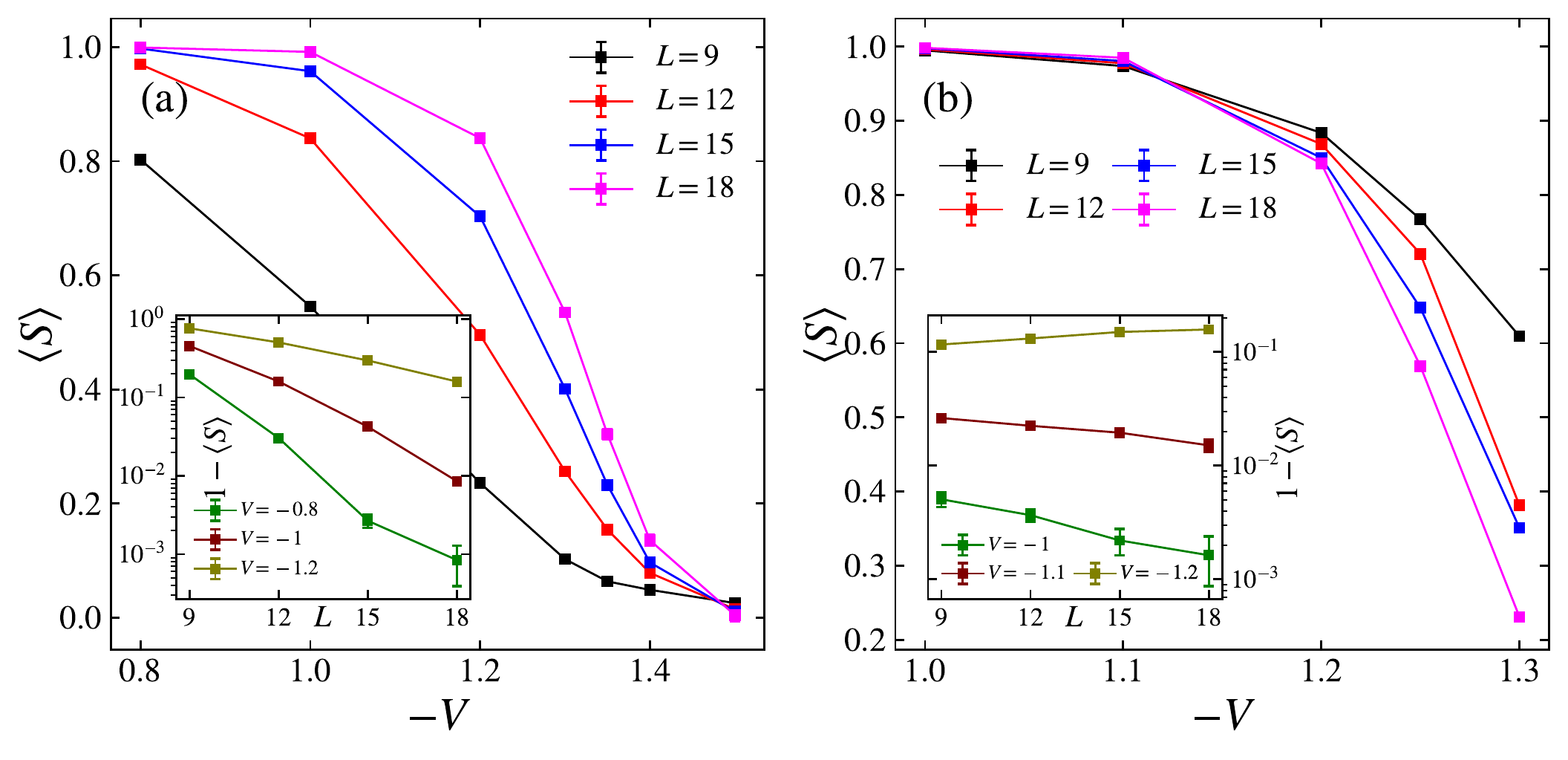}
\caption{(a) The results of average sign $\avg{S}$ in zero-temperature QMC of the spinless fermionic model with attractive interaction $V<0$. The calculation is performed for linear system size $L=9,12,15,18$. The projector parameter is fixed $\Theta = 60$.  (b) The results of average sign  $\avg{S}$ in finite-temperature QMC of the same model as in (a). The calculation is performed for $L=9,12,15,18$. The inverse temperature $\beta$ scales with $L$ as $\beta = 3L$. The insets explicitly show the results of average sign $\avg{S}$ versus $L$ for several values of $V$ in the weak interaction regime. }
\label{Fig3}
\end{figure}

Results of $\avg{S}$ in both finite-temperature and ground-state QMC simulations are shown in \Fig{Fig2}.
Intriguingly, with increasing system size $L$, the average sign $\avg{S}$ displays two distinct behaviors in the regimes of small and large $V$. For zero-temperature QMC simulations, as shown in \Fig{Fig2}(a), the value of $\avg{S}$ increases and approaches to $1$ with increasing system size $L$ in the weak interacting regime ($V<V_*$), indicating that the sign problem asymptotically vanishes in the thermodynamic limit. 
The results of average sign $\avg{S}$ as function of system size $L$ are shown in the insets of \Fig{Fig2}, clearly showing the increasing behaviour of $\avg{S}$ with $L$ in the weak interaction regime.
However, when the interaction is relatively strong ($V>V_*$), $\avg{S}$ decays exponentially as a function of $L$, consistent with the conventional scaling behaviour of sign problem. 
For finite-temperature QMC simulations, as presented in \Fig{Fig2}(b) where inverse temperature $\beta$ scales linearly with system size $L$, the results also provide compelling evidence of asymptotically free behaviour of sign problem in the weak interaction regime.  
Thus, our QMC simulations of the model with repulsive interaction clearly shows the ASF behaviour when the interaction is relatively weak. Note that the sign transition point at $V=V_*$ between two distinct scaling behaviours in the sign problem is close to physical quantum phase transition point $V=V_c\approx 1.355$ \cite{Li2015NJP,Wang2014NJP}.

{\bf Attractive spinless honeycomb model:} For the spinless fermion honeycomb model in \Eq{Model1} with attractive interaction ($V<0$), its QMC simulation is sign-problematic for any known HS transformations.
This model was shown to exhibit pair density wave (PDW) \cite{Jian2015PRL, Kivelson2020ARCMP} ordering in certain range of attractive interaction and also exotic quantum critical point with emergent supersymmetry \cite{Jian2015PRL, Jian2017PRL, Grover2014Science, Li2018ScienceAdvances, Li2017PRL, Yang2008PRL}.
Here, we investigate the scaling behavior of sign problem in this sign-problematic model for both weak and strong attractive interaction.
Again, we do the HS transformation in the density channel
and perform simulations of the model both at zero temperature and at finite temperature.
Surprisingly, in a large regime of weak interactions, the sign problem is asymptotically free, implying the accurate ground-state properties can be achieved reliably by large-scale QMC simulations although the model is not strictly sign free. 

For zero-temperature simulations of the model, the results explicitly reveal that in the regime $|V|\leq 1.2$ the average sign increases and approaches to one with increasing $L$, implying asymptotic sign free behavior in this model, as shown in \Fig{Fig3}(a). As $|V|> 1.2$, the average sign vanishes asymptotically with increasing system size.
This behaviour qualitatively holds in the results of finite-temperature QMC simulation, as shown in \Fig{Fig3}(b), further corroborating the existence of asymptotic sign free behavior in attractive spinless honeycomb model.
Hence, although sign problem is not strictly solved in spinless model \Eq{Model1} with $V<0$, it is still feasible to access the ground-state properties of this model with large system size by QMC simulation in a large regime of interactions and to possibly fathom intriguing physics such as pair-density-wave superconductivity and emergent SUSY, which is left to future study. 

{\bf Renormalization-group-like flow of sign:}
The numerical results above provide compelling evidences that various interacting Dirac-fermion models can exhibit a new type of sign behaviour, dubbed as asymptotic sign free.
Consequently, the sign problem can feature two qualitatively distinct scaling behaviours, depending on the models under study as well as the HS transformations employed in QMC simulations.
A natural question is why there exist two types of distinct behaviours in the scaling of average sign with system size.

Here we propose a RG-like explanation.
For massless Dirac fermion systems in 2D, weak short-range four-fermion interactions are irrelevant in RG, thus rendering the non-interacting limit ($U=0$) a stable fixed point model. For $U=0$, it is obvious that it is sign free with $\avg{S}=1$.
The schematic RG flow of interactions for the Dirac-fermion system is given in \Fig{Fig1}(a).
There exists a critical point $U=U_c$ separating two distinct phases: the Dirac semimetal phase in which interactions are irrelevant and the broken-symmetry phase in which interactions are relevant.
Analogous to the RG flow of interactions as the length scale is increased, we schematically plot the flow of average sign as the system size is increased, as shown in \Fig{Fig1}(b). There are also two distinct behaviours of sign problem separated by $U=U_*$: asymptotic sign free regime and conventional sign-problematic regime.
For asymptotic sign free regime ($U<U_*$), as unveiled in this paper, the interaction is irrelevant and flows to the non-interacting limit; similarly the average sign also flows to $\avg{S}=1$, implying asymptotic sign free.
Conversely, for sufficiently strong interactions, the interaction is relevant and the average sign $\avg{S}$ flows to the point $\avg{S}=0$ as the system size is increased, namely conventional sign-problematic regime.  

\begin{figure}[t]
\includegraphics[scale=0.52]{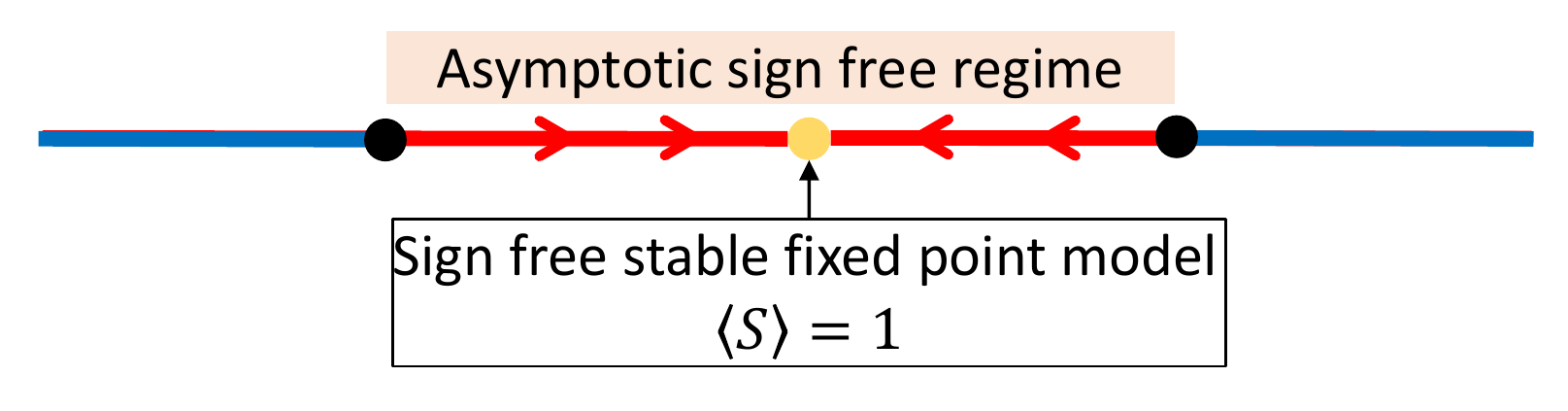}
\caption{The schematic representation of the flow of sign around a stable fixed-point-model which is sign free. We conjecture that around a sign free stable-fixed-point model there exists a finite regime which is asymptotic sign free.}
\label{Fig5}
\end{figure}

The results obtained in the above Dirac fermion models strongly imply that the existence of a sign-free stable fixed point model is essential for the emergence of asymptotic sign free behaviour in QMC simulations.
From this perspective, we propose a conjecture based on our numerical observations and arguments from RG: for an interacting model, if there exists a stable fixed point that is sign-free, a finite asymptotic sign free regime should emerge around this sign free stable fixed point model, as schematically illustrated in \Fig{Fig5}.
%A rigorous proof of it, if any, remains unknown so far, which is left for future study. 
We emphasize that despite the emergence of asymptotic sign free regime, the physical phase transition point is in general not identical to the sign transition point although they can be close to each other. %which depends on the algorithms utilized in the simulation, for example, the choice of decoupling channel in H-S transformation.
A related conjecture is the following: for lattice models featuring generic Fermi surfaces in two or higher dimensions, the non-interacting limit are unstable against infinitesimal attractive interactions; consequently asymptotic sign free regime should not appear around the non-interacting model with Fermi surfaces. This conjecture is numerically confirmed by our calculation of average sign for doped spinless fermion model with relatively weak interaction, as discussed in the Supplementary Materials, which clearly shows that sign average decays exponentially with system size, consistent with the conventional behaviour of sign problem.

{\bf Discussions and concluding remarks:}  %Here we would like to point out the differences and novelties of the asymptotic sign free behaviour found in the present paper in contrast with recent works discussing various special behaviors of sign problem. 
%It was shown that for certain special models the average sign in QMC simulations displays algebraic decay with system size instead of the conventional exponential decay \cite{Zhang2022PRB}.
%Note that for the asymptotic sign free behavior discussed in this paper the average sign actually {\it increases} with system size and approaches to one in the limit of infinite system size.
Recently Ref. \cite{Mondaini2022Science} showed numerical evidences that sign problem in QMC can be most severe close to physical quantum critical points in certain interacting fermion models. 
The present study shows that the sign can be asymptotic free, even close to the quantum critical point. Moreover, the sign transition point $U=U_*$ is in general different from the physical quantum critical point $U=U_c$ although they can be close to each other. Intrinsic relation between nominal critical exponent $\nu_*$ of sign transition and $\nu$ of physical phase transition remains elusive and is left for future study.

In conclusion, we showed convincing evidences for the first time that sign problem can disappear asymptotically with increasing system size, totally beyond the conventional understanding that the average sign decays exponentially with system size.
Moreover, we proposed a RG-like picture to heuristically understand sign problem in interacting models and build the possible correspondence between physical RG flow and the scaling behaviour of sign problem. %In the RG-like picture, the scaling behavior of sign problem may provides a new tool to explore different phases of matters.
Even when sign problem appears in QMC simulations of a given model, it remains feasible to explore quantum correlation physics at low temperature and large system sizes if the model is asymptotic sign free.
Consequently, we believe that our finding in this paper might open a new and promising way to investigate intriguing correlation physics such as high-temperature superconductivity and emergent supersymmetry through asymptotic sign free simulations.

\textit{Acknowledgement}: We would like to thank Shi-Xin Zhang, Yuan Wan and Yuan-Yao He for helpful discussions. This work is supported in part by the NSFC under Grant No. 11825404 (ZQW and HY), the MOSTC Grants No. 2018YFA0305604 and No. 2021YFA1400100 (HY), the CAS Strategic Priority Research Program under Grant
No. XDB28000000 (HY), and the
start-up grant of IOP-CAS (ZXL).

\bibliography{Signproblem-bib}

\begin{widetext}
\section{Supplementary Materials}

\renewcommand{\theequation}{S\arabic{equation}}
\setcounter{equation}{0}
\renewcommand{\thefigure}{S\arabic{figure}}
\setcounter{figure}{0}
\renewcommand{\thetable}{S\arabic{table}}
\setcounter{table}{0}

\subsection{A. The details of quantum Monte-Carlo simulation}
Determinant quantum Monte-Carlo is commonly used to study the finite-temperature or zero-temperature properties of interacting fermionic systems. For finite-temperature algorithm, people deal with quantum partition function of the model: $Z = \Tr{e^{-\beta \hat{H}}}$, where $\hat{H}$ is the Hamiltonian under consideration and $\beta$ is inverse temperature. The ensemble average of the observable is obtained by the QMC simulation: $\avg{O}=\frac{\Tr{[e^{-\beta \hat{H}}\hat{O}}]}{\Tr{e^{-\beta\hat{H}}}}$.  For zero-temperature algorithm, the ground-state expectation value of given observable is evaluated by:
\bea
\langle\hat{O}\rangle=\frac{\langle\psi_G|\hat{O}|\psi_G\rangle}{\langle\psi_G|\psi_G\rangle} =\lim_{\Theta\ra\infty}\frac{\langle\psi_T|e^{-\frac{\Theta}{2} H}\hat{O}e^{-\frac{\Theta}{2} H}|\psi_T\rangle}{\expectation{\psi_T}{e^{- \Theta H}}{\psi_T}}
\eea
where $\Theta$ is projector parameter, $\ket{\psi_G}$ is the ground-state wave function of the Hamiltonian $\hat{H}$, and $\ket{\psi_T}$ is a slater-determinant trial wave function that has non-zero overlap with $\ket{\psi_G}$.  The Determinant QMC involves two main steps as following: (1) Trotter-Suzuki decomposition, which discretizes inverse temperature $\beta$ (projector parameter $\Theta$) into small imaginary-time slices: $\Delta_\tau = \frac{\beta}{N_\tau}$($\Delta_\tau=\frac{\Theta}{N_\tau}$) in finite-temperature (zero-temperature) algorithm.   (2) Hubbard-Stratonovic (HS) transformation, under which four-fermions terms in Hamiltonian are decoupled to the non-interacting bilinear-fermion operators coupled to classical auxiliary fields. As we discuss below, the appearance of sign problem depends on the decoupling channel of H-S transformation. In our study, we fix $\Delta_\tau=0.1$ in the study of spinful Hubbard and $\Delta_\tau = 0.05$ in the study of spinless t-V model, which is sufficient to guarantee the convergence of results. In finite-temperature simulation, we scale $\beta$ linearly with system size $L$. In zero-temperature simulation, we have checked that $\Theta=60$ is sufficiently large to access the accurate ground-state physical observables in spinless t-V for the system sizes under consideration in our simulation. Hence, we fix $\Theta=60$ and calculate average sign for different $L$ in the simulation on spinless t-V models.

For spinful Hubbard model, we implement an unconventional H-S transformation by decoupling the Hubbard interaction differently between terms on A sublattices and B sublattices. We decouple the Hubbard interaction in the spin channel as following:
\bea
e^{-\Delta \tau U(n_{i\uparrow}-\frac 1 2)(n_{i\downarrow}-\frac 1 2)}=\frac{1}{2} e^{-U\Delta \tau/4}\sum_{s_i =\pm 1}e^{\lambda s_i c^\dagger_i \vec{\sigma} \cdot \vec{n}_i c_i}
\eea
where $c^\dagger_i = (c^\dagger_{i\uparrow},c^\dagger_{i\downarrow})$, $\cosh \lambda = e^{\frac{U\Delta_\tau}{2}}$ and $s_i$ is auxiliary field defined on site $i$. Specifically, we set $\vec{n}_i = \hat{z}$ for $i$ in A sublattice and $\vec{n}_i = \vec{n} = (\sin\theta,0,\cos\theta)$ on B sublattice. As $\theta=0$, this HS transformation becomes the conventional one and the model is sign free at half filling. When $\theta \neq 0$, the sign problem appears and the severity of sign problem depends on the value of $\theta$. In the maintext, we present the results of the simulation by choosing $\theta = \frac{\pi}{2}$.

For repulsive spinless t-V model, the model is sign-free at half filling if the interaction is decoupled in the hopping channel, as introduced in previous works. The interaction can be decoupled in hopping channel as:
\bea
 e^{-\Delta_\tau V (n_i-\frac{1}{2})(n_j-\frac{1}{2})} = \frac{1}{2}e^{-V\Delta_\tau/4} \sum_{s_{ij}=\pm 1} e^{\lambda s_{ij} (c^\dagger_i c_j + h.c.)}
\eea
where $\cosh \lambda = e^{\frac{V\Delta_\tau}{2}}$. The absence of sign problem is identified from various perspectives, including Majorana time-reversal symmetry \cite{Li2015PRB,Li2016PRL}, split orthogonal group \cite{Wang2015PRL} and Majorana reflection positivity \cite{Xiang2016PRL}. For attractive interaction, the sign problem of the model cannot be eliminated in any known algorithms of QMC simulation. Hence the ground-state properties of the spinless t-V model with attractive remains elusive. In our study, to investigate the properties of sign problem, we perform H-S transformation of spinless t-V model in a sign-problematic channel, namely density channel. In this decoupling channel, the t-V model is sign-problematic for both repulsive and attractive interaction, even at half filling. For repulsive interaction $V>0$:
\bea
 e^{-\Delta_\tau V (n_i-\frac{1}{2})(n_j-\frac{1}{2})} = \frac{1}{2}e^{-V\Delta_\tau/4} \sum_{s_{ij}=\pm 1} e^{\lambda s_{ij} (n_i - n_j)}
\eea
where $\cosh \lambda = e^{\frac{V\Delta_\tau}{2}}$ and $s_{ij}$ is auxiliary field defined on bond $ij$. For attractive interaction $V<0$:
\bea
e^{-\Delta_\tau V (n_i-\frac{1}{2})(n_j-\frac{1}{2})} = \frac{1}{2}e^{-V\Delta_\tau/4} \sum_{s_{ij}=\pm 1} e^{\lambda s_{ij} (n_i + n_j-\frac{1}{2})}
\eea
where $\cosh \lambda = e^{\frac{-V\Delta_\tau}{2}}$ and $s_{ij}$ is auxiliary field defined on bond $ij$. Sign problems exist for both repulsive and attractive interactions in this decoupling channel. We perform simulation on spinless t-V model in this decoupling channel, and demonstrate the existence of asymptotic sign-free regime at half filling for both repulsive and attractive interactions.

\begin{figure}[t]
\includegraphics[scale=0.5]{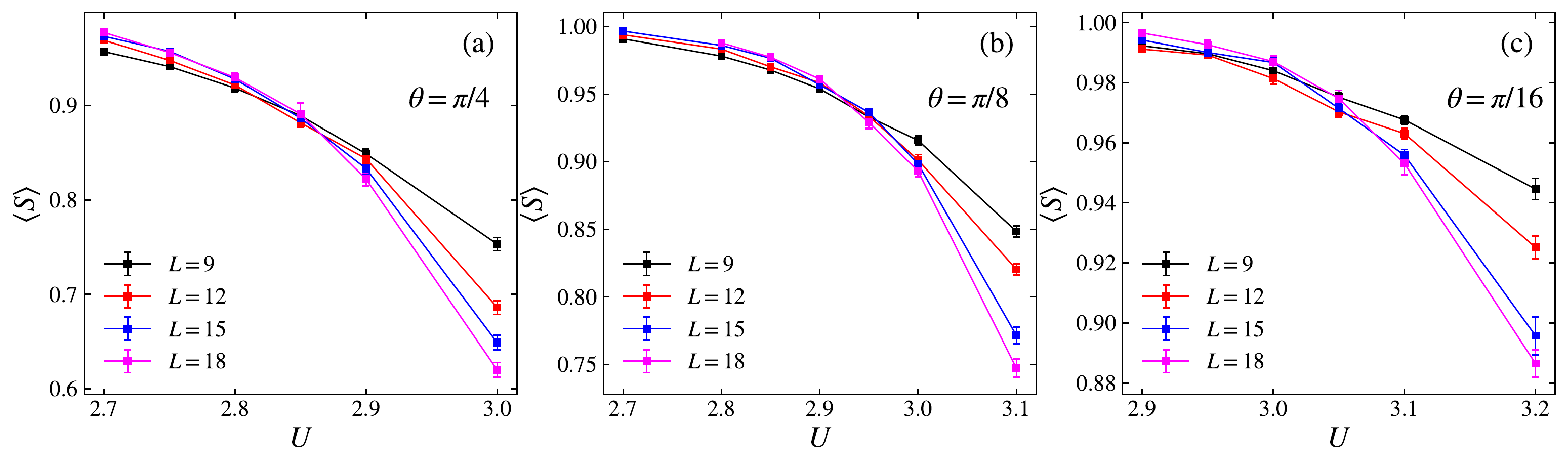}
\caption{The results of finite temperature QMC for the average sign $\left<S\right>$ as a function of interaction strength $U$ in honeycomb Hubbard model at half filling using partially rotated HS transformations with (a) $\theta = \pi/4$; (b) $\theta = \pi/8$; (c) $\theta = \pi/16$. The crossing region of $\left<S\right>-U$ curves has a slight shift when $\theta$ is changing. The calculation is performed for linear system size $L=9,12,15,18$. The inverse temperature $\beta$ scales with linear system size as $\beta = 2 L$. The simulation parameter $\Delta \tau=0.1$.}
\label{FigS1}
\end{figure}

\subsection{B. Additional results of sign problem in honeycomb Hubbard model}
In this section, we demonstrate that the ASF region with weak interaction exists using different HS transformations in honeycomb Hubbard model at half filling. We show the results of average sign $\left<S\right>$ as a function of interaction strength U using partially rotated HS transformations with rotation parameter $\theta=\pi/4,\pi/8,\pi/16$ in \Fig{FigS1}. In all these three cases, the average sign show different scaling behaviors between strong and weak interaction region separated by a critical $U_*$, indicting asymptotic sign free region exist for different HS transformations. The critical interaction $U_*$ has a shift when changing parameter $\theta$ ($U_*\sim 2.85,2.9,3.0$ for $\theta=\pi/4,\pi/8,\pi/16$, respectively). Since its value can change for different HS transformation, we do not imply any direct relationship between $U_*$ and the physical critical point $U_c$. Nevertheless, the asymptotic sign free behaviors still exist in the weak interaction region using different HS transformation in honeycomb Hubbard model at half filling.

\begin{figure}[b]
\includegraphics[scale=0.45]{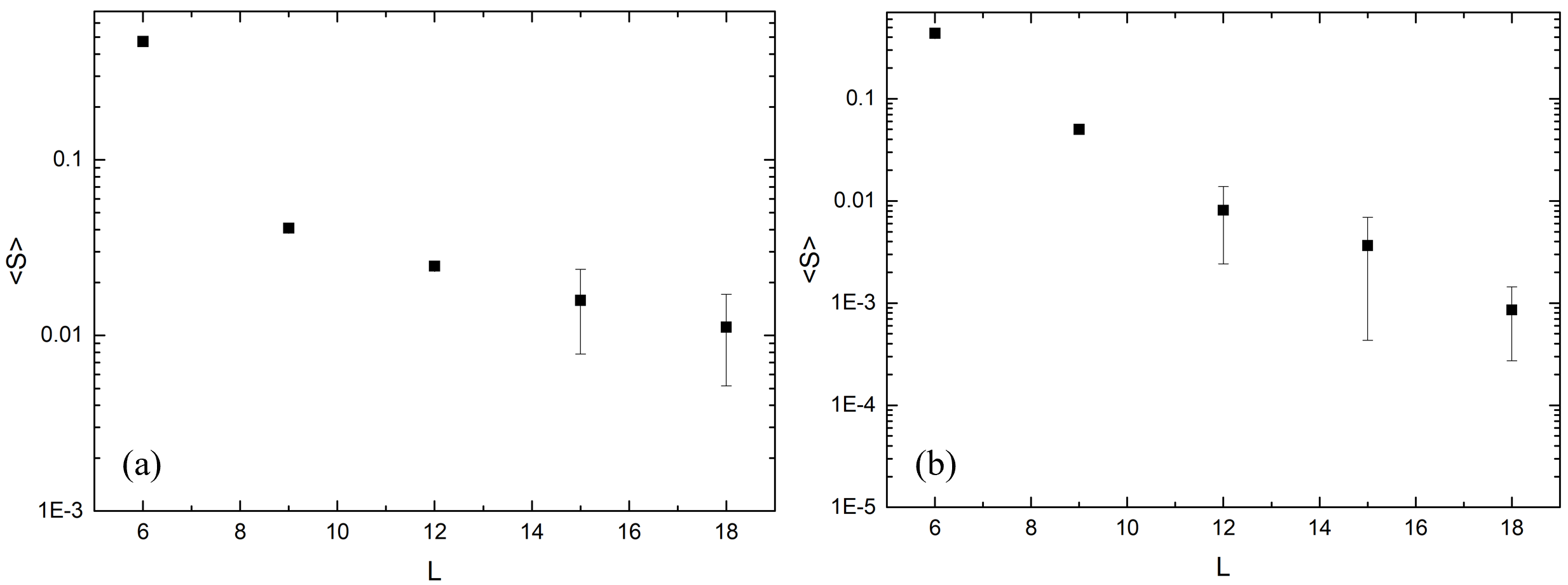}
\caption{The finite-temperature QMC results of average sign for doped repulsive (a) and attractive (b) spinless t-V model. The interaction strength is fixed $V=0.8$. The doping level is $p=0.04$. The inverse temperature $\beta$ scales with linear system size $L$ as $\beta=3L$. }
\label{FigS3}
\end{figure}

\subsection{C. The sign problem in doped spinless t-V model with Fermi surfaces}
In the main text, we have shown that for both repulsive and attractive spinless t-V model at half filling, the models are asymptotic sign-free in a large parameter region. In this section, we discuss the behaviour of sign problem occurring in spinless t-V model away from half filling. We perform finite-temperature QMC simulation and scale inverse temperature $\beta$ with linear system size $L$ as $\beta=3L$. In finite-temperature QMC simulation, doping is achieved by tuning chemical potential. In \Fig{FigS3}, we present the results of sign averages at a typical interaction strength $|V|=0.8$ deep in the asymptotic sign-free regime at half filling. The results of average sign $\avg{S}$ as a function of linear system size $L$ for repulsive and attractive interactions are shown in \Fig{FigS3} (a) and (b), respectively, clearly indicating the values of $\avg{S}$ decay with system size in both cases. As system size increases, the $\avg{S}$ displays exponentially decaying behaviour consistent with the conventional scaling behaviour of sign problem in QMC. Taking other interaction strengths and doping levels, the average sign exhibits qualitatively same behaviours, decaying with increasing system size and lowering temperature.

\end{widetext}

\end{document}